\documentclass{Interspeech}

\interspeechcameraready

\title{Egocentric Speaker Classification in Child-Adult Dyadic Interactions: From Sensing to Computational Modeling}

\author[affiliation={1}]{Tiantian}{Feng}
\author[affiliation={1}]{Anfeng}{Xu}
\author[affiliation={1}]{Xuan}{Shi}
\author[affiliation={2}]{Somer}{Bishop}
\author[affiliation={1}]{Shrikanth}{Narayanan}

\affiliation{Signal Analysis and Interpretation Laboratory}{University of Southern California}{USA}
\affiliation{Weill Institute for Neurosciences}{University of California, San Francisco}{USA}
\email{tiantiaf@usc.edu}

\usepackage{comment}
\usepackage{multirow}
\usepackage{xcolor}
\usepackage{booktabs}
\usepackage{tikz}
\usepackage{subcaption}
\usepackage{stfloats}
\usepackage{tipa}
\usepackage{tabularx}
\usepackage{comment}
\begin{document}

\maketitle

\begin{abstract}
Autism spectrum disorder (ASD) is a neurodevelopmental condition characterized by challenges in social communication, repetitive behavior, and sensory processing.
One important research area in ASD is evaluating children's behavioral changes over time during treatment. The standard protocol with this objective is BOSCC, which involves dyadic interactions between a child and clinicians performing a pre-defined set of activities. A fundamental aspect of understanding children's behavior in these interactions is automatic speech understanding, particularly identifying who speaks and when. Conventional methods in this area heavily rely on audio recorded from a spectator perspective, and there is limited research on egocentric speech modeling. Here, we experiment with wearable sensors to perform speech sampling in BOSCC interviews from an egocentric perspective. Our findings highlight the potential of egocentric speech collection and pre-training to improve speaker classification accuracy.
\end{abstract}

\section{Introduction}
\label{sec:intro}

Autism spectrum disorder (ASD) \cite{lord2018autism, lord2020autism} is a neurocognitive developmental disorder that describes individuals with impairments in social communication and interactions along with repetitive and restrictive behaviors. Children with ASD may face challenges in acquiring language skills and communicating with others, which can increase the risk of developing mental disorders such as anxiety and depression. Recent studies show that early treatment of ASD can lead to substantial improvement in a child's development \cite{daniolou2022efficacy}. Evaluating the behavior changes over time during treatment involves clinicians conducting child speech and behavior analysis. Language samples are typically acquired through established clinical protocols involving semi-structured dyadic interactions between a child and clinicians to evaluate their communicative abilities and behaviors. One prominent protocol is \textit{Brief Observation of Social Communication Change (BOSCC)}~\cite{grzadzinski2016measuring} for tracking changes in social and communicative skills during treatment. These clinical observations aim to elicit spontaneous responses from children under different circumstances to obtain the effectiveness of treatment.

% Even though ASD may be diagnosed at any age, many critical signs and symptoms of ASD can be observed early in life. Specifically, recent studies report that 1 in 36 children in the USA are diagnosed with ASD \cite{maenner2023prevalence}.

Automatic understanding of speech samples involving child interactions in this clinically relevant context creates numerous opportunities, including neurocognitive disorder assessment and treatment. One fundamental step of processing speech signals in these interactions requires a robust and precise computational method to classify child and adult labels of speech segments. The prior studies in \cite{xu2024audio, lahiri2023robust} have shown the importance of recognizing child and adult speaker labels to achieve automated spoken language assessment. Combined with the recent developments in speech foundation models \cite{bommasani2021opportunities}, substantial progress has been made in the last few years in automatic child speech understanding \cite{fan24b_interspeech, xu23e_interspeech, blockmedin24_interspeech, zhang24d_interspeech, demopoulos24_interspeech, graave24_interspeech, li24j_interspeech, johnson2023recanvo}, particularly in child-adult speaker classification \cite{koluguri2020meta, xu2024audio, li2021analysis}.

\begin{figure}[t]
	\centering
	\includegraphics[width=0.95\linewidth]{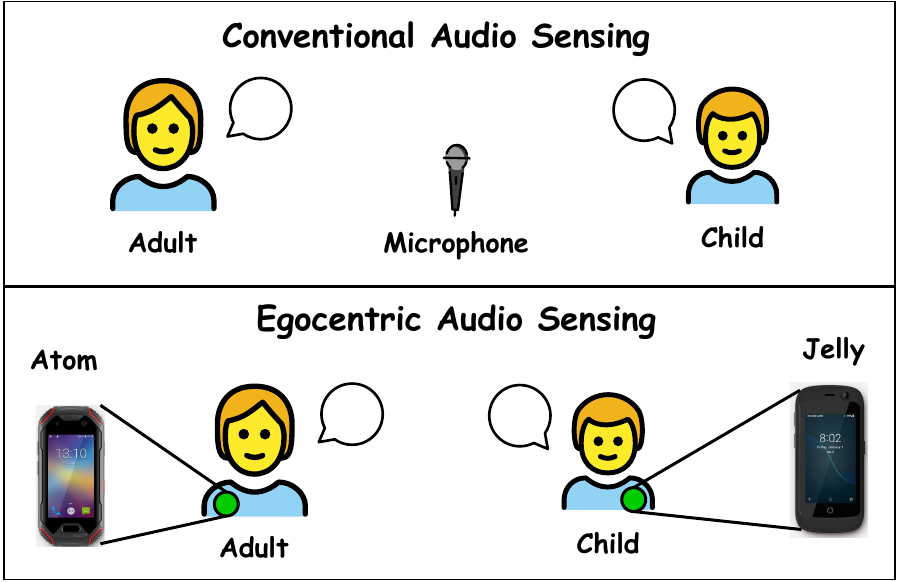}
    \vspace{-1mm}
    \caption{Comparisons between conventional and egocentric audio sampling methods for child-adult dyadic interactions. In the egocentric audio sensing setup, both adults and children wore wearable audio sensors clipped close to their mouths.}
    \vspace{-5mm}
    \label{fig:experiments}
\end{figure}

While these efforts have led to great promises, most existing works focus on using speech samples recorded from a ``spectator'' perspective, such as microphones placed in a stationary environment. As a result, these previous findings cover only a narrow spectrum of speech sensing and understanding involving child and adult dyadic interactions, such as far-field speech processing. Moreover, developing robust speaker classification models from the far-field speech setup with stationary mics is challenging. Instead, in this work, we focused on continuous speech data from the ``first-person" perspective, where the recordings were tied to the hearing and uttering of the child or the adult interacting with the environment. Specifically, our study setup involving child-adult dyadic interactions is closely related to the massive-scale audio-visual data presented in the Ego4D \cite{grauman2022ego4d}. Specifically, our contributions include:

\begin{itemize}[leftmargin=*]
  \item  We conducted one of the earliest studies that collect egocentric speech signals using a wearable audio sensing device in child-adult interactions in a clinically relevant context.
  
  \item Instead of relying on existing pre-trained models for child-adult speaker recognition, the proposed work integrates a pre-training stage using the egocentric speech from the Ego4D.
  
  \item We perform extensive fine-tuning experiments on the dataset collected from egocentric speech samples involving child-adult interactions following BOSCC protocol. Our results demonstrate the effectiveness of pre-training for egocentric sensing in improving child-adult speech classification. 
  
  % However, diarization experiments indicate that egocentric audio pre-training benefits more for adult speech.
  
\end{itemize}

\begin{figure}[t]
	\centering
	\includegraphics[width=0.95\linewidth]{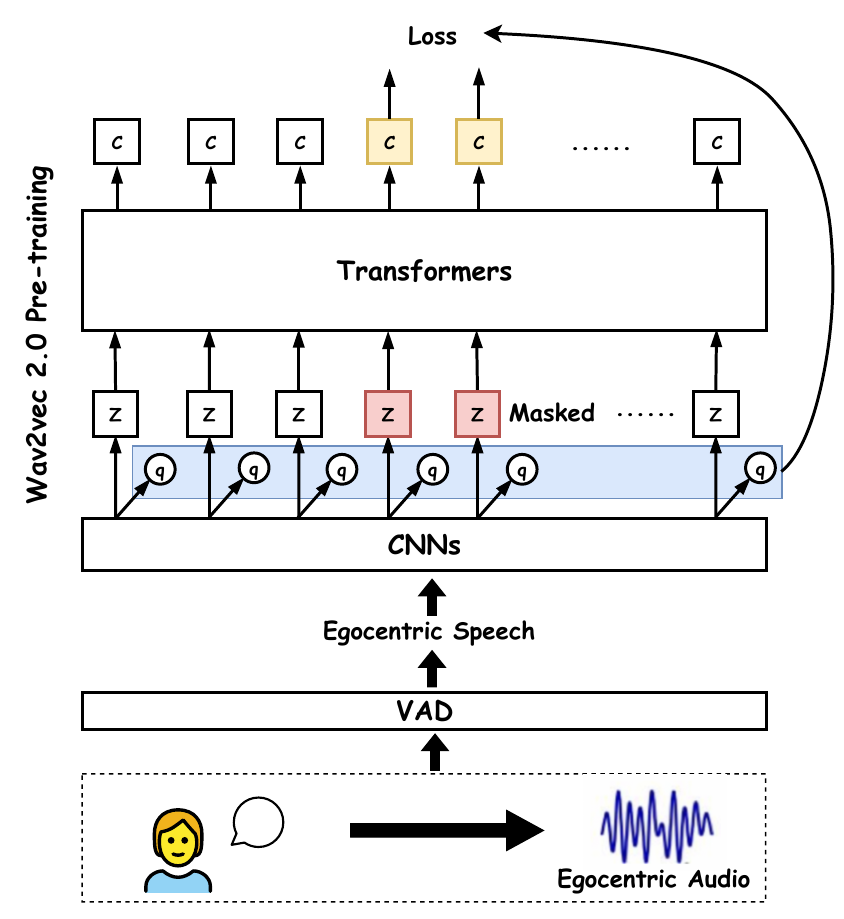}
    \vspace{-2mm}
    \caption{Our proposed pretraining framework. The Ego4D audio samples were first processed by VAD.}
    \label{fig:pretraining}
    \vspace{-3.5mm}
\end{figure}

\section{Egocentric Audio Sensing}
\label{sec:method}

\noindent \textbf{Sensing technology} In this study, we propose to collect egocentric speech from child-adult interactions using wearable sensors. Although audio sensing technologies have evolved significantly in the last two decades, many systems remain bulky, expensive, and lack scalability. Recent technological advances in wearable sensors have enabled the emergence of applications for capturing social communications from an egocentric perspective in naturalistic settings \cite{liang2023dynamic, feng2018tiles, rachuri2010emotionsense, hebbar2023behavioral}. For example, by using the miniature mics on these devices, a digital recorder \cite{Mehl2001, Liue1601185, Lena, feng2018tiles} can overcome many drawbacks of the traditional systems, capturing day-long audio data centered around a person. In this work, we prototype the experiments following a previously proposed wearable solution called TILES Audio Recorder (TAR) \cite{feng2018tiles}. We would highlight that this wearable solution applies to both controlled and naturalistic settings.

% \cite{feng2018tiles}

\noindent \textbf{Study Design} The primary motivation for us to adopt TAR is its open-source nature and compatibility with various form factors, including lightweight and budget-friendly smartphones like the Jelly and Atom \footnote{https://www.unihertz.com/}. We modify the original functionality in TAR to collect raw audio rather than low-level acoustic descriptors during child-adult dyadic interactions. Specifically, these child-adult interactions are conducted following the BOSCC protocol. In the experiment, both the child and the examiner wore TAR devices throughout the sessions. The child used the lightweight Jelly version for comfort, while the examiner wore the Atom, as shown in Figure~\ref{fig:experiments}. It is worth noting that we observed a better audio recording quality using Atom sensors than Jelly sensors. 

% \section{Data Collection}
% \label{sec:data_collection}

\vspace{0.5mm}
\noindent \textbf{Data Collection} The data collection was a part of the extension to the study described in \cite{byrne2022behavioral}. Children who participated in this study were instructed to wear the Jelly version of the TAR during their visit to conduct BOSCC interviews. The data includes ten unique BOSCC sessions from 10 children between the ages of 2-7 years old. Particularly, among all children, six are male. Given that this study is an early effort to pilot egocentric audio sensing involving child-adult interactions, this cohort of data collection includes only three children with ASD.

\section{Egocentric Speech Modeling}
\label{sec:pretraining}

\subsection{Pre-training The Backbone Model}
Previous works have reported promising child-adult speaker classification results leveraging the speech foundation models. However, most existing speech foundation models are trained with speech recordings in a non-egocentric perspective. To improve the ability of these pre-trained speech models for egocentric speaker prediction, we perform a pre-training stage on a large-scale egocentric speech dataset called Ego4D.

\vspace{0.5mm}
\noindent \textbf{Ego4D Dataset} The Ego4D dataset is a popular multimodal dataset that contains egocentric audio, videos, and IMUs from diverse geographic coverage, environments, and populations. The dataset includes over 3,500 hours of video data from 932 unique participants in 9 countries. 

\vspace{0.5mm}
\noindent \textbf{Voice Activity Detection} As most audio data from the Ego4D are unlabeled continuous naturalistic recordings, there exists a substantial amount of silence or noise in these data. Therefore, we decide to preprocess the audio to obtain speech-only segments for the pre-training. Specifically, we leverage the voice activity detection model from pyannote~\cite{bredin2021end} to filter out speech segments from continuous egocentric recordings.

\vspace{0.5mm}
\noindent \textbf{Wav2vec 2.0 Pre-training} In this work, we primarily investigate the wav2vec 2.0 pre-training  \cite{baevski2020wav2vec} on Ego4D speech samples. As shown in Figure~\ref{fig:pretraining}, the wav2vec 2.0 model consists of convolutional encoders followed by transformers. The speech data $X$ is first mapped to latent feature $Z$, and the transformers model $Z$ to the contextualized representation $C$. Moreover, the latent feature $Z$ is discretized to $Q$ with a quantization module to form the contrastive objective. It is worth noting that the quantization module leverages the Gumbel Softmax operation to generate $Q$. The primary objective in wav2vec 2.0 pre-training is the contrastive object to distinguish the true quantized latent speech and its associated contextual representations after masking. More details regarding wav2vec 2.0 pre-training can be found in \cite{baevski2020wav2vec}. We initialized the weights with the wav2vec 2.0-base model weights before further pre-training with speech samples filtered from the Ego4D dataset.

\begin{figure}[t]
	\centering
	\includegraphics[width=0.95\linewidth]{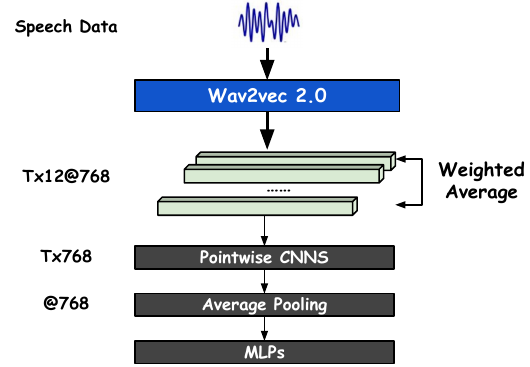}
    \caption{Proposed fine-tuning architecture.}
    \label{fig:fintune}
    \vspace{-4mm}
\end{figure}

\begin{figure}[t]
	\centering
	\includegraphics[width=0.9\linewidth]{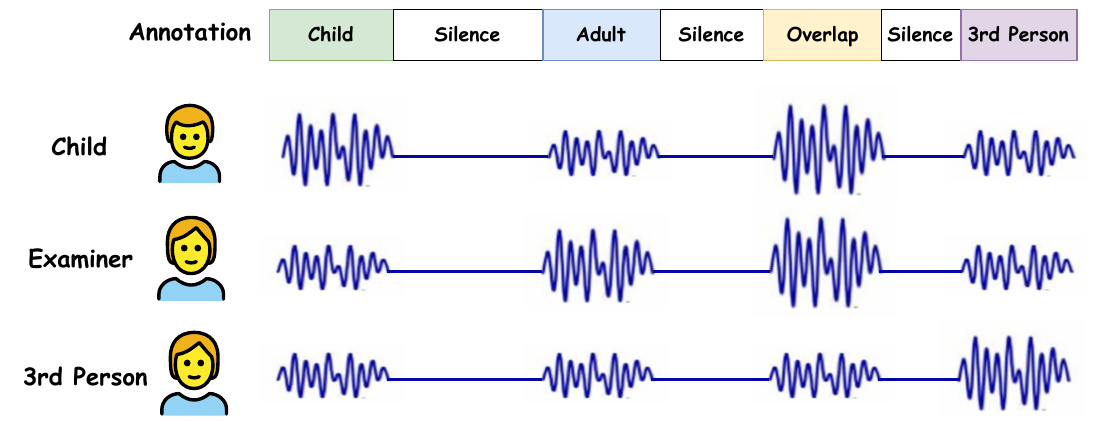}
    \vspace{-2mm}
    \caption{An example of annotated speaker labels.}
    \label{fig:annotation}
    \vspace{-4mm}
\end{figure}

\subsection{Speaker Modeling}
\label{sec:finetune}
Once we complete the pre-training on Ego4D data, we perform the fine-tuning following the architecture from the previous work \cite{xu23e_interspeech}. Specifically, we apply a linear layer to combine hidden outputs from all layers. The weighted output is then processed through a set of pointwise 1D convolutional layers, an average pooling layer, and MLPs for prediction. We apply the ReLu activation functions within convolutional and MLP layers. The model architecture is demonstrated in Figure~\ref{fig:fintune}.

\section{Experimental Details}
\label{sec:experiments}

\subsection{Data Annotation}
As our paper focuses on classifying child and adult speech, we conducted manual annotations using four labels: child, adult, 3rd parties, and overlap. Due to IRB restrictions, the annotations were carried out by researchers within the same research group who received proper research training certificates. In particular, background noise was excluded from the annotations. Third-party speakers could include other family members, such as siblings, another parent, or toddlers. Our modeling experiments focus exclusively on speech samples labeled as child or adult. In total, our annotation includes 4,240 unique speech samples across 10 BOSCC sessions, with 2,709 examiner (adult) speech samples and 1,531 child speech samples.

\subsection{Modeling Details}

\noindent \textbf{Ego4D Pre-training:} The model weights of the pyannote VAD model were obtained from Huggingface for processing the Ego4D. Since this study was conducted using American English, we decided to filter out Ego4D audio data with the recording location within the United States. Moreover, those audios with a recording duration of less than $5$ minutes were also excluded from pre-training. After VAD, there were approximately 75,000 speech utterances for pre-training. We initialize the pre-training model with the wav2vec 2.0 weights from Huggingface\footnote{https://huggingface.co/facebook/wav2vec2-base}. Specifically, we choose a masking ratio of $50\%$, a learning rate of 0.0001, and a total training epoch of 40.

\vspace{1mm}
\noindent \textbf{Speaker Classification:} We apply a learning rate in \{2e-4, 5e-4\} for speaker classification training, with the CNN layers having an output channel size of 256. We perform the classification fine-tuning using two unique backbones: the wav2vec 2.0 base (W2V) and the Ego4D wav2vec 2.0 base (Ego4D W2V).  
% For the diarization experiment, we adopt the opensource diarization framework described in \cite{xu2024data} and use a learning rate of 1e-3.
We perform 5-fold cross-validation with two unique BOSSC sessions as a test fold and report the average performance across all folds. The reported metrics include accuracy, macro-F1, recall, and specificity for the classification task. In the classification task, adult and child labels are 0 and 1, respectively. 
% , and diarization error rate (DER) for the diarization task

\section{Results}
\label{sec:experiments}

\subsection{Egocentric Child-Adult Speaker Modeling}

We begin by exploring whether pre-training the backbone model using egocentric speech samples improves child-adult speaker classification in an egocentric recording setup, as shown in Table~\ref{tab:baseline_result}. 
In this experiment, the training data includes mono-channel speech samples from devices worn by both the child and the examiner. 
Compared with results reported in \cite{xu23e_interspeech} (Macro F1 = 0.798), sampling audio with an egocentric setup helps to improve speaker classification.
When tested on speech samples recorded from different sources (child or examiner), the findings suggest that pre-training the backbone model with egocentric speech samples consistently improves child-adult speaker classification. 
Specifically, we notice that the performance gains mainly originate from increased specificity, implying better adult speech classification. Given that most Ego4D speech samples come from adults, this highlights the benefit of pre-training on adult egocentric speech. 
It also points to the need for more egocentric speech samples from children to improve child speaker classification.

\begin{table}[t]
    % \footnotesize
    \caption{Child-adult speaker classification between two backbones: wav2vec 2.0 base (W2V) and a version further pre-trained with Ego4D speech samples (E4D W2V). The adult and child speaker labels are 0 and 1, respectively. The training data includes speech from both the child and the examiner.}
    
    \resizebox{0.48\textwidth}{!}{
        \begin{tabular}{lccccc}
            \toprule
            \multirow{1}{*}{\textbf{Test Set}} & \multirow{1}{*}{\textbf{Backbone}} & \textbf{Acc} & \textbf{F1} & \textbf{Recall} & \textbf{Specificity} \\
            \cmidrule(lr){1-2} \cmidrule(lr){3-6} 
            
            \multirow{2}{*}{Jelly-Child} & W2V & $84.0$ & $0.801$ & $\mathbf{77.2}$ & $87.8$ \rule{0pt}{1.65ex} \\
            
            & Ego4D W2V & $\mathbf{85.4}$ & $\mathbf{0.812}$ & $74.2$ & $\mathbf{90.7}$ \rule{0pt}{1.65ex} \\
    
            \midrule
            
            \multirow{2}{*}{Atom-Exam} & W2V & $84.8$ & $0.810$ & $\mathbf{81.3}$ & $87.6$ \rule{0pt}{1.65ex} \\
            & Ego4D W2V & $\mathbf{87.5}$ & $\mathbf{0.842}$ & $80.9$ & $\mathbf{90.8}$ \rule{0pt}{1.65ex} \\
            
            \bottomrule
        \end{tabular}
    }
    
    \vspace{-3mm}
    \label{tab:baseline_result}
\end{table}

\subsection{Impact of Training Recording Resource}

We then compare the performances using a single source of recordings for training as shown in Table~\ref{tab:sources}. Specifically, we explore the training with domain matching and domain mismatching. Specifically, domain matching indicates that the training and test have the same recording source, e.g., train and test both with recordings from the device the child is wearing (Child/Child). Instead, the domain mismatching represents the training and testing with recordings from different devices.

% train the speech samples from either children or the examiner and test the models with the same or different recording sources.

\vspace{0.5mm}
\noindent \textbf{Domain Matching (Child/Child and Examiner/Examiner):} Similar to the results presented earlier, we find that pre-training with the egocentric speech samples can consistently improve the child-adult speaker classification regardless of the source of training speech samples. Moreover, we note a higher child-adult classification performance when using the recording source from the examiner, confirming the better audio quality from the device worn by the examiner. We also observe that performance gains mainly come from improved adult speech classification.

\vspace{0.5mm}
\noindent \textbf{Domain Mismatching (Child/Examiner and Examiner/Child):} Even when training and testing speech samples with varying sources of recording, pre-training backbone models with egocentric speech samples still help child-adult speaker modeling in these scenarios. However, we notice a consistent improvement in recall scores (child speech classification) when training and testing are used with speech samples from different recording sources.

% Moreover, this experimental setup can be regarded as a proxy for the conventional recording setup, where the testing can be regarded as far-field speech However, testing with different recording sources consistently underperforms with testing using the recording source.

\begin{table}
    \footnotesize
    \caption{Child-adult speaker classification results using different recording sources as training data. The adult and child speaker labels are 0 and 1, respectively.}
    \centering
    \vspace{-1mm}
    \resizebox{0.49\textwidth}{!}{
        \begin{tabular}{lcccccc}
    
            \toprule
            % \multirow{1}{*}{\textbf{Train}} & \multirow{1}{*}{\textbf{Test}} & \multirow{1}{*}{\textbf{Backbone}} & \multicolumn{4}{c}{\textbf{Classification}} & \textbf{Diarization} \\ 

            \multirow{1}{*}{\textbf{Train}} & \multirow{1}{*}{\textbf{Test}} & \multirow{1}{*}{\textbf{Backbone}} &  \textbf{Acc} & \textbf{F1} & \textbf{Recall} & \textbf{Specificity} \\
            \cmidrule(lr){1-3} \cmidrule(lr){4-7} 
    
            \multirow{4}{*}{Child} & \multirow{2}{*}{Child} & W2V & $81.6$ & $0.766$ & $\mathbf{75.7}$ & $84.3$ \rule{0pt}{1.65ex} \\
            & & Ego4D W2V & $\mathbf{83.9}$ & $\mathbf{0.790}$ & $74.0$ & $\mathbf{86.7}$ \rule{0pt}{1.65ex} \\

            \cmidrule(lr){2-7}
             & \multirow{2}{*}{Exam} & W2V & $76.4$ & $0.684$ & $51.1$ & $\mathbf{88.3}$ \rule{0pt}{1.65ex} \\
            & & Ego4D W2V & $\mathbf{78.8}$ & $\mathbf{0.716}$ & $\mathbf{55.8}$ & $88.0$ \rule{0pt}{1.65ex} \\
            
            \midrule
            \multirow{4}{*}{Exam} & \multirow{2}{*}{Child} & W2V & $67.1$ & $0.671$ & $78.0$ & $\mathbf{70.0}$ \rule{0pt}{1.65ex} \\
            
            & & Ego4D W2V & $\mathbf{70.4}$ & $\mathbf{0.674}$ & $\mathbf{79.8}$ & $69.3$ \rule{0pt}{1.65ex} \\
            \cmidrule(lr){2-7}
            
            & \multirow{2}{*}{Exam} & W2V & $88.6$ & $0.861$ & $\mathbf{85.3}$ & $90.9$ \rule{0pt}{1.65ex} \\
            & & Ego4D W2V & $\mathbf{89.6}$ & $\mathbf{0.873}$ & $83.9$ & $\mathbf{93.4}$ \rule{0pt}{1.65ex} \\

            \bottomrule
        \end{tabular}
    }
    \vspace{-3mm}
    \label{tab:sources}
\end{table}

\begin{figure}[t] {
    \centering
    \begin{tikzpicture}

        \node[draw=none,fill=none] at (0, 3.0){\includegraphics[width=\linewidth]{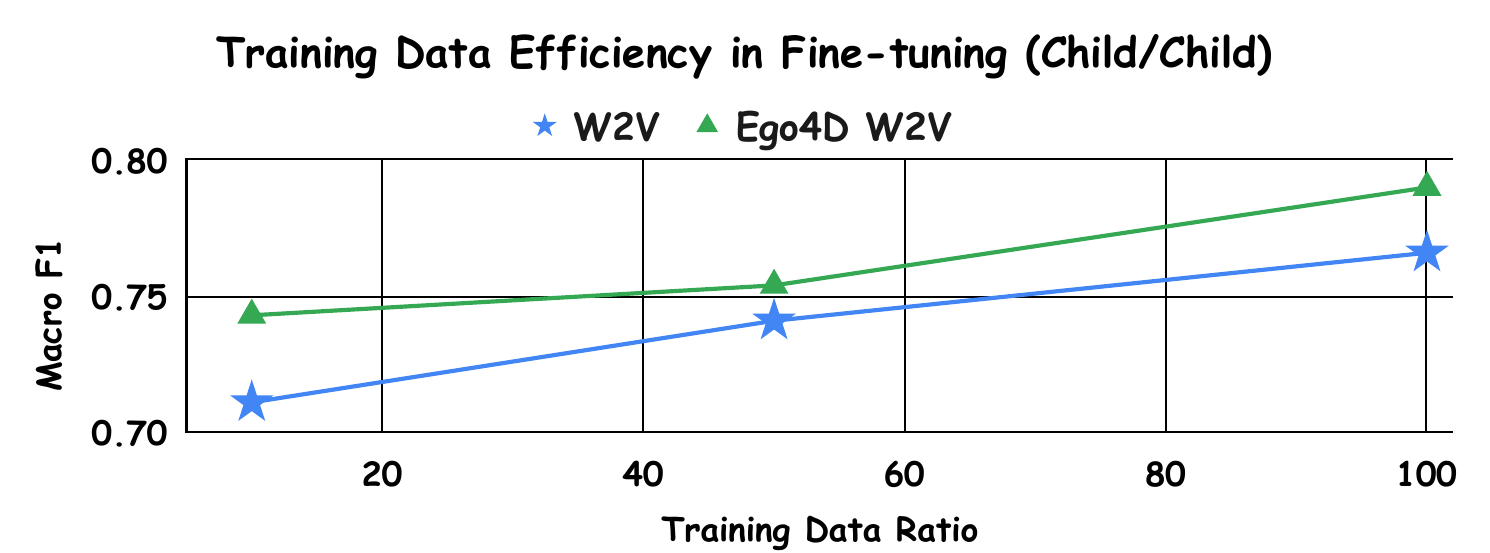}};

        \node[draw=none,fill=none] at (0,0){\includegraphics[width=\linewidth]{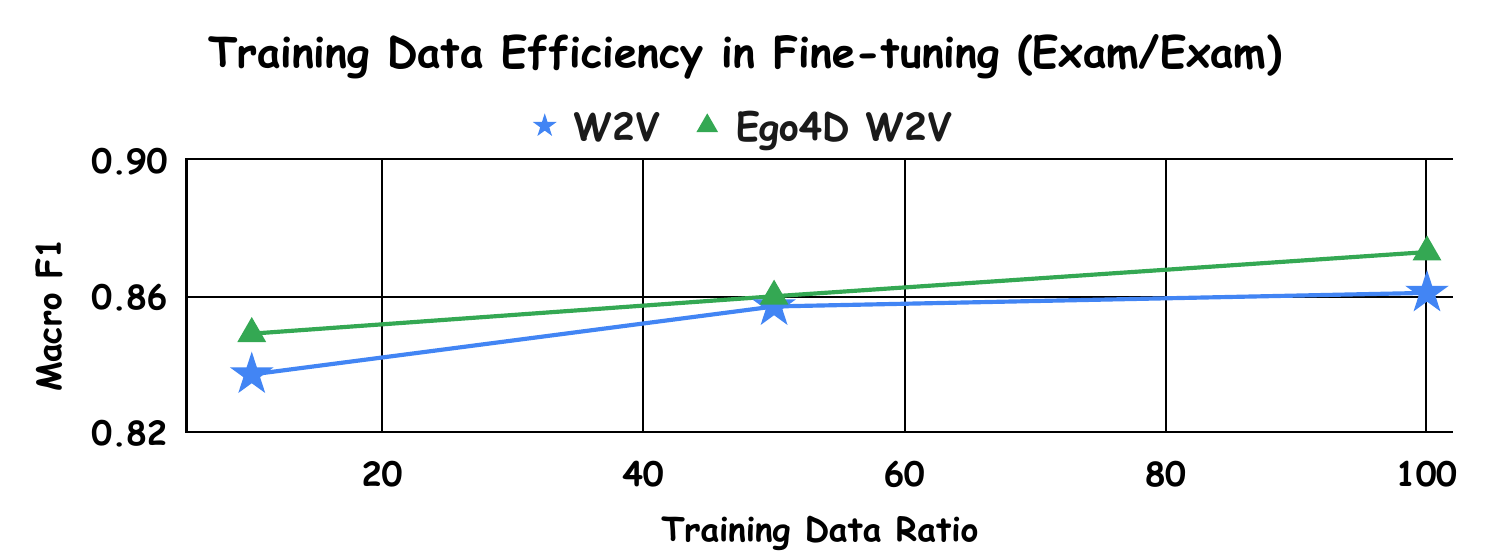}};

    \end{tikzpicture}
    \vspace{-3mm}
    \caption{Performance at different training data ratios.}
    \label{fig:efficiency}
    \vspace{-3.5mm}
} \end{figure}

\subsection{Training Data Efficiency}
Apart from leveraging complete annotated data for fine-tuning, we explore the efficiency of training data in child-adult speaker modeling. Due to the limited space, we choose to only present speaker classification results. Precisely, we compare two backbones in fine-tuning the child-adult speaker classification using $10\%, 50\%, 100\%$ of training samples. 
Results in Figure~\ref{fig:efficiency} demonstrate that Ego4D W2V consistently outperforms W2V across all training data ratios. These findings further support the effectiveness of integrating pre-training stages on egocentric speech samples to improve child-adult speaker classification sampled in the egocentric setup.

\begin{figure}[t] {
    \centering
    
    \begin{tikzpicture}

        \node[draw=none,fill=none] at (0, 2.6){\includegraphics[width=\linewidth]{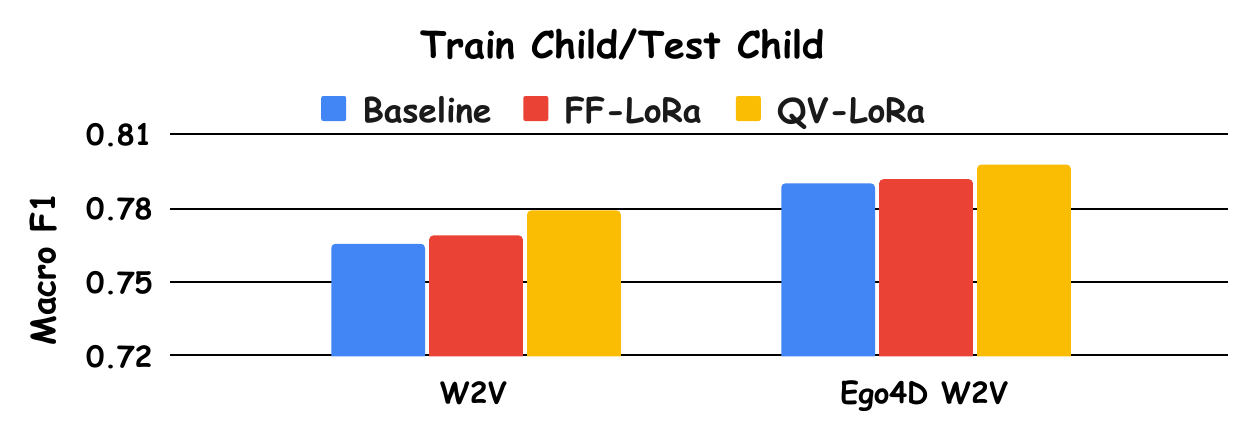}};

        \node[draw=none,fill=none] at (0,0){\includegraphics[width=\linewidth]{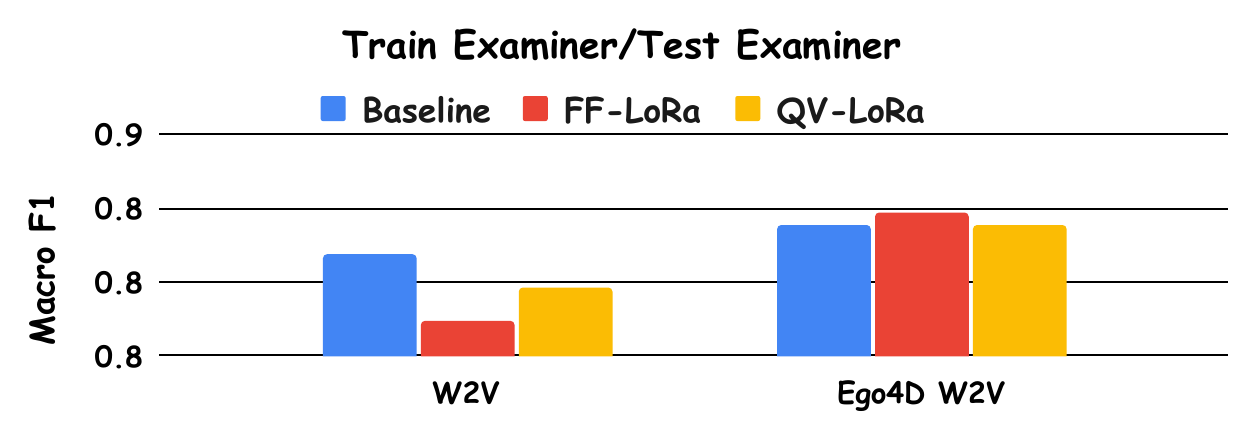}};
        
    \end{tikzpicture}
    \vspace{-7mm}
    \caption{Performance comparisons between PEFT and without PEFT. FF-LoRA implements LoRA in feedforward layers, and QV-LoRA uses LoRA for query and value matrices.}
    \label{fig:lora}
} \end{figure}

\subsection{Generalizability to PEFT}

In addition to fine-tuning with the pre-trained speech embeddings, we aim to investigate whether the backbone models can be adapted to other training approaches, such as parameter-efficient fine-tuning (PEFT) \cite{pmlr-v97-houlsby19a}. To investigate this, we experimented with one popular PEFT approach, LoRA (Low-rank Adaptation) \cite{hu2021lora}. The primary idea behind LoRA is to fine-tune the low-rank matrices to approximate the model updates associated with linear layers. We implement FF-LoRA (feedforward layers) and QV-LoRA (query and value matrices) in this experiment. The PEFT results in Figure~\ref{fig:lora} show that LoRA consistently improves child-adult classification for Ego4D W2V backbones. At the same time, there is a decrease in performance using LoRA in W2V backbones with the examiner device recording as training and testing data. We also observe that FF-LoRA performs better than QV-LoRA in our experiment.

\subsection{Dual Channel Modeling}

Apart from modeling the child-adult speaker with the mono-channel input, we experimented with a concatenated input that combines speech samples from both devices. Similar to the PEFT experiment, we only explore the classification task. We would highlight that this modeling approach is similar to the dual-channel method proposed in \cite{li24j_interspeech}. Overall, the comparisons in Tab~\ref{tab:dual} demonstrate that input with dual channels further improves child-adult speaker classification compared to the best-performing mono-channel approach. 
Compared with the previous work with conventional recording setup \cite{xu23e_interspeech}, dual-channel egocentric speech with Ego4D W2V backbone yields a 0.08 improvement in macro F1.
Moreover, we can also observe a substantial performance increase in specificity using the Ego4D W2V backbone. This further implies the improved adult speaker classification that is associated with pre-training egocentric adult speech samples from the Ego4D dataset.

\begin{table}
    \footnotesize
    \caption{Child-adult speaker classification with dual channel modeling. The adult and child speaker labels are 0 and 1, respectively. The training samples include a two-channel speech from the child and the examiner.}
    \centering

    \resizebox{0.47\textwidth}{!}{
        \begin{tabular}{lcccccc}
    
            \toprule
            \textbf{Input} & \textbf{Backbone} & \textbf{Acc} & \textbf{F1} & \textbf{Recall} & \textbf{Specificity} \\ 
            \midrule
    
            % Mono \cite{xu23e_interspeech} & W2V & - & $0.798$ & - & - \rule{0pt}{1.65ex} \\
            
            Mono (Best) & Ego4D W2V & $87.5$ & $0.842$ & $80.9$ & $90.8$ \rule{0pt}{1.65ex} \\
            % \cmidrule(lr){2-6}
            \midrule
            
             \multirow{2}{*}{Dual} & W2V & $87.9$ & $0.855$ & $\mathbf{84.7}$ & $91.1$ \rule{0pt}{1.65ex} \\
            
            & Ego4D W2V & $\mathbf{89.7}$ & $\mathbf{0.873}$ & $79.6$ & $\mathbf{95.3}$ \rule{0pt}{1.65ex} \\

            \bottomrule
    
        \end{tabular}
    }
    
    \vspace{-3mm}
    \label{tab:dual}
\end{table}

\section{Conclusion and Future Works}

In this work, we explore the use of wearable audio-sensing to sample egocentric speech samples from child-adult interactions in clinical observations. We design the egocentric audio sensing experiments leveraging a wearable audio sensor tested in large-scale longitudinal studies in naturalistic settings. Moreover, we curate the data by annotating the speaker labels based on our prior work in \cite{xu23e_interspeech}.
On the other hand, we propose a modeling framework that begins with pre-training the backbone model with speech samples from the Ego4D dataset. Our extensive experiments show that speech samples collected in the egocentric setup can help pre-trained speech models to better classify child and adult speech. Moreover, pre-training the backbone model with egocentric speech samples can further improve the child-adult speaker classification. Our future work includes comparing speaker diarization with previous child-adult speaker diarization models \cite{xu2024data} and comparing with other speech foundation models.

\section{Acknowledgements}
We gratefully acknowledge support from the Simons Foundation (award number: SFI-AR-HUMAN-00004115-03, 655054). Some of our images are mixed from OpenMoji.

\bibliographystyle{IEEEtran}
\bibliography{refs}

\end{document}